\documentclass[preprint,12pt]{elsarticle}
\newcommand{\pip}{\ensuremath{\pi^+\,}}
\newcommand{\pim}{\ensuremath{\pi^-\,}}
\newcommand{\piz}{\ensuremath{\pi^0\,}}
\newcommand{\etapippimpiz}{\ensuremath{\eta\rightarrow\pip\pim\piz}}

\def\ifm#1{\relax\ifmmode#1\else$#1$\fi}

  \def\x{\ifm{\times}}
  
\def\pt#1,#2,{\ifm{#1\x10^{#2}}}

\def\pim{\ifm{\pi^-}}  \def\pip{\ifm{\pi^+}}

\makeatletter
\newdimen\z@ \z@=0pt 
\newskip\z@skip \z@skip=0pt plus0pt minus0pt
\def\m@th{\mathsurround=\z@}
\def\ialign{\everycr{}\tabskip\z@skip\halign} 
\def\eqalign#1{\null\,\vcenter{\openup\jot\m@th
  \ialign{\strut\hfil$\displaystyle{##}$&$\displaystyle{{}##}$\hfil
      \crcr#1\crcr}}\,}
\makeatother

\newcommand{\affuni}[2]{Dipartimento di Fisica dell'Universit\`a #1, #2, Italy.}

\newcommand{\affinfnm}[2]{INFN Sezione di #2, #2, Italy.}
\newcommand{\affinfnn}[2]{INFN Sezione di #1, #2, Italy.}

\journal{Physics Letters B}
\begin{document}
  
  \begin{frontmatter}
    \title{Measurement of the $\eta\rightarrow 3\pi^{0} $ slope parameter $\alpha$ with the KLOE detector}

    \author[Na,infnNa]{F.~Ambrosino\corref{cor1}} \ead{Fabio.Ambrosino@na.infn.it}
    \author[Frascati]{A.~Antonelli}
    \author[Frascati]{M.~Antonelli}
    \author[Roma2,infnRoma2]{F.~Archilli}
    \author[Mainz]{P.~Beltrame}
    \author[Frascati]{G.~Bencivenni}
    \author[Roma1,infnRoma1]{C.~Bini}
    \author[Frascati]{C.~Bloise}
    \author[Roma3,infnRoma3]{S.~Bocchetta}
    \author[Frascati]{F.~Bossi}
    \author[infnRoma3]{P.~Branchini}
    \author[Frascati]{P.~Campana}
    \author[Frascati]{G.~Capon}
    \author[Na]{T.~Capussela\corref{cor1}} \ead{Tiziana.Capussela@na.infn.it}
\author[Roma3,infnRoma3]{F.~Ceradini}
\author[Frascati]{P.~Ciambrone}
\author[Frascati]{E.~De~Lucia}
\author[Roma1,infnRoma1]{A.~De~Santis}
\author[Frascati]{P.~De~Simone}
\author[Roma1,infnRoma1]{G.~De~Zorzi}
\author[Mainz]{A.~Denig}
\author[Roma1,infnRoma1]{A.~Di~Domenico}
\author[infnNa]{C.~Di~Donato}
\author[Roma3,infnRoma3]{B.~Di~Micco}
\author[Frascati]{M.~Dreucci}
\author[Frascati]{G.~Felici}
\author[Frascati]{A.~Ferrari}
\author[Roma1,infnRoma1]{S.~Fiore}
\author[Roma1,infnRoma1]{P.~Franzini}
\author[Frascati]{C.~Gatti}
\author[Roma1,infnRoma1]{P.~Gauzzi}
\author[Frascati]{S.~Giovannella}
\author[Frascati]{M.~Jacewicz}
\author[Karlsruhe]{W.~Kluge}
\author[Moscow]{V.~Kulikov}
\author[Frascati,StonyBrook]{J.~Lee-Franzini}
\author[Frascati,Energ]{M.~Martini}
\author[Na,infnNa]{P.~Massarotti}
\author[Na,infnNa]{S.~Meola}
\author[Frascati]{S.~Miscetti}
\author[Frascati]{M.~Moulson}
\author[Mainz]{S.~M\"uller}
\author[Frascati]{F.~Murtas}
\author[Na,infnNa]{M.~Napolitano}
\author[Roma3,infnRoma3]{F.~Nguyen}
\author[Frascati]{M.~Palutan}
\author[infnRoma3]{A.~Passeri}
\author[Frascati,Energ]{V.~Patera}
\author[Na,infnNa]{F.~Perfetto\corref{cor1}} \ead{Francesco.Perfetto@na.infn.it}
\author[Frascati]{P.~Santangelo}
\author[Frascati]{B.~Sciascia}
\author[Frascati,Energ]{A.~Sciubba}
\author[Frascati]{T.~Spadaro}
\author[Roma3,infnRoma3]{C.~Taccini}
\author[infnRoma3]{L.~Tortora}
\author[infnRoma1]{P.~Valente}
\author[Frascati]{G.~Venanzoni}
\author[Frascati,Energ]{R.Versaci}
\author[Frascati,Beijing]{G.~Xu}
\address[Frascati]{Laboratori Nazionali di Frascati dell'INFN, 
Frascati, Italy.}
\address[Karlsruhe]{Institut f\"ur Experimentelle Kernphysik, 
Universit\"at Karlsruhe, Germany.}
\address[Mainz]{Institut f\"ur Kernphysik, 
Johannes Gutenberg - Universit\"at Mainz, Germany.}
\address[Na]{Dipartimento di Scienze Fisiche dell'Universit\`a 
``Federico II'', Napoli, Italy}
\address[infnNa]{INFN Sezione di Napoli, Napoli, Italy}
\address[Energ]{Dipartimento di Energetica dell'Universit\`a 
``La Sapienza'', Roma, Italy.}
\address[Roma1]{\affuni{``La Sapienza''}{Roma}}
\address[infnRoma1]{\affinfnm{``La Sapienza''}{Roma}}
\address[Roma2]{\affuni{``Tor Vergata''}{Roma}}
\address[infnRoma2]{\affinfnn{Roma Tor Vergata}{Roma}}
\address[Roma3]{\affuni{``Roma Tre''}{Roma}}
\address[infnRoma3]{\affinfnn{Roma Tre}{Roma}}
\address[StonyBrook]{Physics Department, State University of New 
York at Stony Brook, USA.}
\address[Beijing]{Institute of High Energy 
Physics of Academica Sinica,  Beijing, China.}
\address[Moscow]{Institute for Theoretical 
and Experimental Physics, Moscow, Russia.}

\cortext[cor1]{Corresponding authors}

\begin{abstract}
      We present a measurement of the slope parameter $\alpha$ for the $\eta\rightarrow 3\pi^{0}$
      decay, with the KLOE experiment at the DA$\Phi$NE
      $\phi$-factory, based on a background free sample of $\sim$ 17 millions $\eta$ mesons produced in $\phi$ radiative decays.
      By fitting the event density in the Dalitz plot we determine 
      $\alpha = -0.0301 \pm 0.0035\,stat\;_{-0.0035}^{+0.0022}\,syst\,$. The result is in
      agreement with recent measurements from hadro- and photo-production experiments.
   
    \end{abstract}
    
  \end{frontmatter}
  
  \section{Introduction}
  \label{introduzione}
  
\begin{figure}[!htb] 
    \begin{center}
	\includegraphics[width=.5\linewidth]{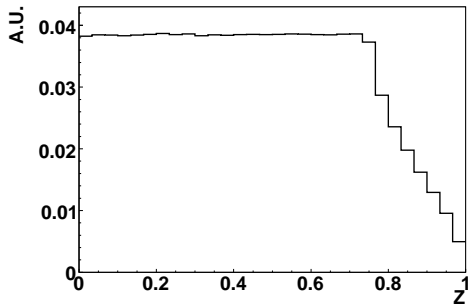}\\
      \caption{Expected $z$ distribution for pure phase space.} 
      \label{fig:Dalitzplot} 
    \end{center}
  \end{figure}

  The decay $\eta\rightarrow 3 \pi$, \pip\pim\piz and 3\piz, though is a major
  decay mode of the $\eta$ meson, violates isospin symmetry. 
  Since contributions from the electromagnetic interaction are 
  strongly suppressed by chiral symmetry~\cite{Sutherland} this decay 
  is mainly due to the isospin breaking part of the QCD Lagrangian:
  \begin{equation}
    {\cal L}_{\not\, I} = -\frac{1}{2}\left(m_{u}-m_{d}\right)
    \left(\bar{u}u-d\bar{d}\right) 
  \end{equation}
  \noindent
  so that in principle it offers a way to determine the mass
  difference of the up-down quarks.
  Moreover, the selection rule $\Delta I =1$ allows us to relate the
  amplitudes for the two decays using isospin symmetry:
  \begin{equation}
    A_{000}\left(s,t,u\right) = A_{+-0}\left(s,t,u\right) +
    A_{+-0}\left(t,u,s\right) + A_{+-0}\left(u,s,t\right) 
    \\
  \end{equation}
  \noindent
  Theoretical predictions for the decay amplitude have been obtained in
  the framework of Chiral Perturbation Theory (ChPT): the low energy
  effective field theory for QCD. Leading order (LO) ChPT 
  predictions~\cite{BijGa02} based on current algebra
  underestimate the $\eta$ decay rates by a factor of $\simeq 4$. One loop (NLO)
  calculations which include the $\pi - \pi$
  rescattering effects~\cite{Gasser_Leut} predict higher rates but
  still below the observed values. Some improvements are obtained by 
  computing unitary corrections~\cite{KWW} to NLO using a dispersion relation for the decay amplitude
  derived by Khuri and Treiman~\cite{KT}. Recently, more 
  advanced calculations have become available. In reference~\cite{Borasoy} the
  authors use U(3) ChPT, in combination with a coupled channels method,
  and treat final state interactions by means of the Bethe Salpeter
  equation obtaining good agreement with measured decay widths and
  spectral shapes. In Reference~\cite{Bij_Gho} a full NNLO computation is performed showing
  sizable corrections to the NLO result.\\
  The Dalitz plot of a three body decay is described by two
  kinematical variables which for three identical
  particles in the final state, reduce to a single. 
  In the $\eta\rightarrow 3\piz$ decay this variable is chosen by
  convention to be:
  \begin{equation}
    z = \frac{2}{3} \sum_{i=1}^{3} \left (\frac{3E_{i} -
      m_{\eta}}{m_{\eta} - 3m_{\piz}} \right )^{2}, 
    \label{eq:zeta}
  \end{equation}
  \noindent
  where $E_{i}$ denote the energy of the i-th pion in the $\eta$ rest
  frame (CM).
  The variable $z$ lies in the interval $[0-1]$, where $z = 0$
  corresponds to events with
  3 \piz having all the same energy while for $z = 1$ one \piz is
  at rest  and the remaining two are emitted back to back.\\ 
  The decay amplitude is represented at leading order in terms of
  a single quadratic slope parameter $\alpha$:
  \begin{equation}
    \vert A_{000}\left(z\right) \vert^{2} \sim 1 + 2\alpha z.
  \end{equation}
  \noindent
  In case of pure phase space ( i.e. at leading order in ChPT) one has
  $\alpha = 0$ and the $z$ distribution is flat from $z = 0$ to  $z \sim 0.76$
  and then falls to zero at $z = 1$, see Fig.~\ref{fig:Dalitzplot}. 
  \\ 
  Recent measurements of $\alpha$ with $\eta$-mesons produced almost
  at rest in hadro- and photo-production experiments are 
  reported in Table~\ref{tab:exp_theo}.
  \begin{table}[!htb]
    \begin{center}
      \begin{tabular}{||c||c||} 
	\hline
	& {\bf $\alpha$}\\ 
	\hline      
	Crystal Ball (2001)~\cite{C_B01}          & -0.031 $\pm$ 0.004 \\
	CELSIUS WASA~\cite{C_W}                  & -0.026 $\pm$ 0.014 \\
	WASA at COSY~\cite{W_C}                   & -0.027 $\pm$ 0.009 \\
	Crystal Ball at MAMI-B~\cite{C_B09_B}     & -0.032 $\pm$ 0.003
	\\    
	Crystal Ball at MAMI-C~\cite{C_B09}       & -0.032 $\pm$ 0.003 \\    
	\hline
	\hline
	ChPT / LO                                          & 0.000 \\
	ChPT / NLO~\cite{Gasser_Leut}                     & 0.015  \\
	ChPT / NLO + unit. corrections~\cite{KWW}          & -0.014 $\div$ -0.007 \\
	U(3) ChPT + Bethe Salpeter~\cite{Borasoy}          & -0.031 $\pm$ 0.003 \\      
	ChPT / NNLO~\cite{Bij_Gho}                     & 0.013  $\pm$ 0.032\\      
	    \hline
      \end{tabular}
    \end{center}
    \caption{Experimental and theoretical results for the slope parameter $\alpha$.}
    \label{tab:exp_theo}
  \end{table}
  In the same Table are also shown the theoretical
  estimates for $\alpha$ previously described. The
  predicted values show differences - due to large
  cancellations in the amplitude computation - even quoting in some
  case a positive sign for $\alpha$ contrary to the experimental
  evidence. A precise measurement of $\alpha$ therefore
  poses a significant constraint to theoretical models. 
We present a new measurement of $\alpha$ based on a large sample of
  $\eta$ mesons produced in $e^{+} e^{-}$ collisions via the radiative
  decay $\phi\rightarrow\eta\gamma$. \\
  \section{DA$\Phi$NE and KLOE}
  \label{rivelatori}
  Data were collected with the KLOE detector at DA$\Phi$NE~\cite{DAFNE},
  the Frascati $e^{+} e^{-}$ collider, which operates at a
  center of mass energy $W =m_{\phi}\sim 1020$ MeV. 
   The beams collide with a crossing angle of $\pi - 25$ mrad, 
  producing  $\phi$ mesons with a small transverse momentum , $p_{\phi}\sim$ 13 MeV/c.
  The KLOE~\cite{KLOE} detector is inserted in a 0.52 T magnetic field and it consists of a large cylindrical drift chamber (DC), surrounded by a fine
  sampling lead-scintillating fibers electromagnetic calorimeter
  (EMC). 
  The DC~\cite{DC}, 4 m diameter and 3.3 m long, has full stereo
  geometry and operates with a gas mixture of 90\% helium and 10\%
  isobutane.
  Momentum resolution is $\sigma(p_\bot)/p_\bot\leq
  0.4\%$. Position resolution in $r - \phi$ is 150 $\mu$m and $\sigma_{z}\sim$ 2
  mm. Charged tracks vertices are reconstructed with an accuracy of $\sim$ 3 mm.\\
  The EMC~\cite{EMC} is divided into a barrel and two endcaps, and covers 98\% of the solid angle.
  It is segmented into 2440 cells of cross section $\sim 4.4 \times
  4.4$ cm$ ^{2}$ in the plane perpendicular to the fibers. Each cell is read out 
  at both ends by photomultiplier tubes.\\
  Arrival times of particles 
  and space positions of the energy deposits are obtained 
  from the signals collected at the two
  ends; cells close in time and space are grouped into a calorimeter
  cluster.  The cluster energy $E$ is the sum of the cell energies, while
  the cluster time $t$ and its position $r$ are energy weighted
  averages. The energy and time resolutions are respectively $
  \sigma_E/E = 5.7\% / \sqrt{E\ ({\rm GeV})}$
  and $\sigma_t = 57 \;\textrm{ps} / \sqrt{E\ ({\rm GeV})} \oplus 100\;
  \textrm{ps}$. Cluster positions are measured with a resolution of 1.3
  cm in the coordinate transverse to the fibers, and, by timing, of 
$ 1.2\ {\rm cm}/ \sqrt{E\ ({\rm GeV})}$ in the longitudinal coordinate.\\
  The KLOE trigger~\cite{TRIGGER} is based on the coincidence of two  
  energy deposits with $E > 50$ MeV in the barrel and $E > 150$ MeV in the
  endcaps. Moreover, to  reduce the trigger rate due
  to cosmic rays crossing the detector, events with a large energy
  release in the outermost calorimeter planes are vetoed.
  
  \section{Event selection}
  \label{Section:selezione}
\begin{figure}[!htb] 
    \begin{center}
      \includegraphics[width=.5\linewidth]{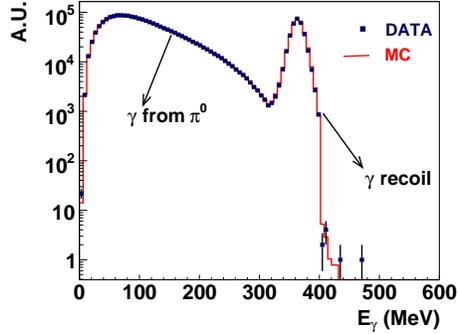}\\
	\caption{ Photon energy spectrum in the laboratory. (Dots: data, histogram: MC).}
      \label{fig:Egamma} 
    \end{center}
  \end{figure}

\begin{figure}[!htb] 
    \begin{center}
      \begin{tabular}{c c}
	\includegraphics[width=.5\linewidth]{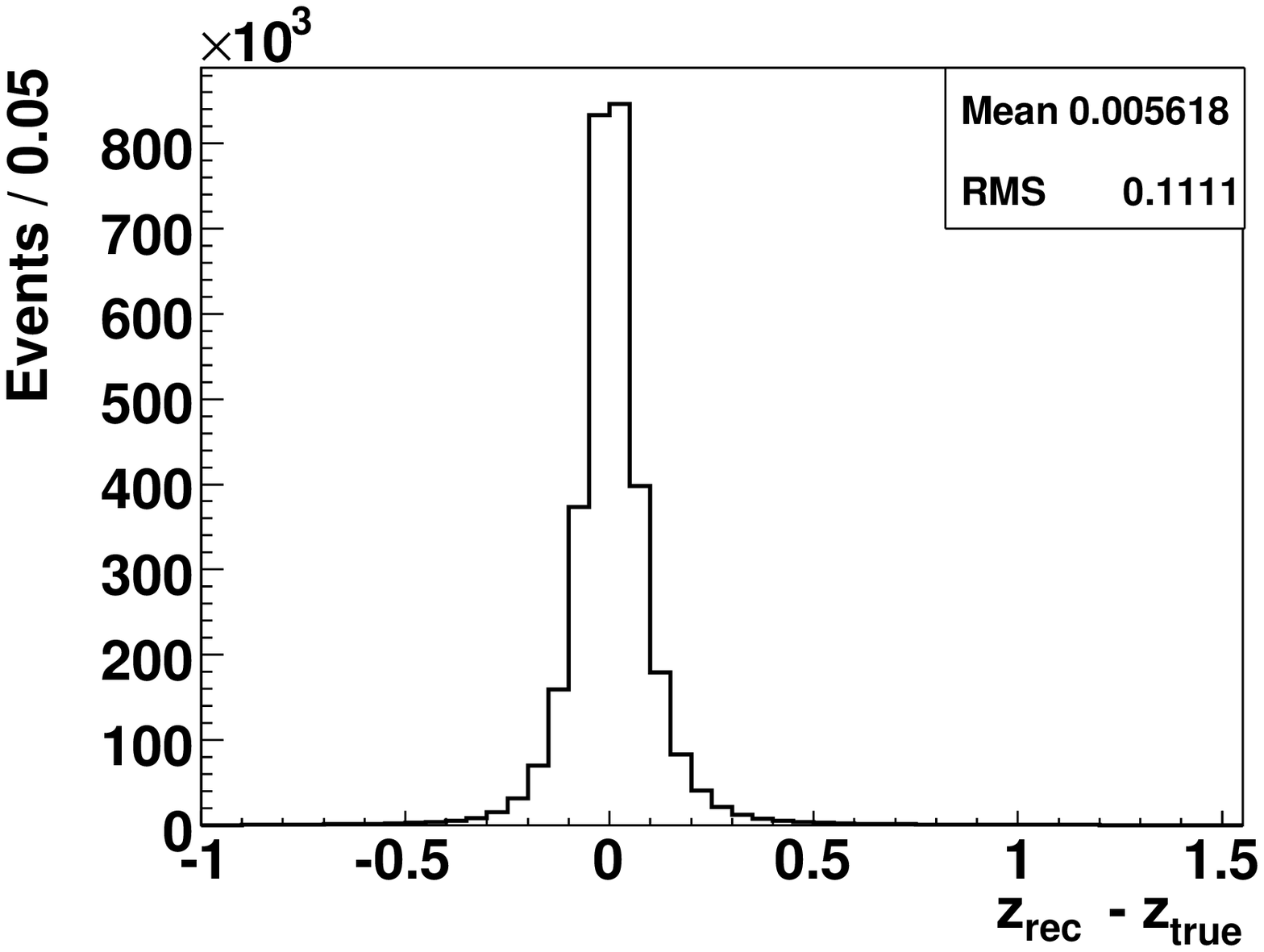}&
	\includegraphics[width=.5\linewidth]{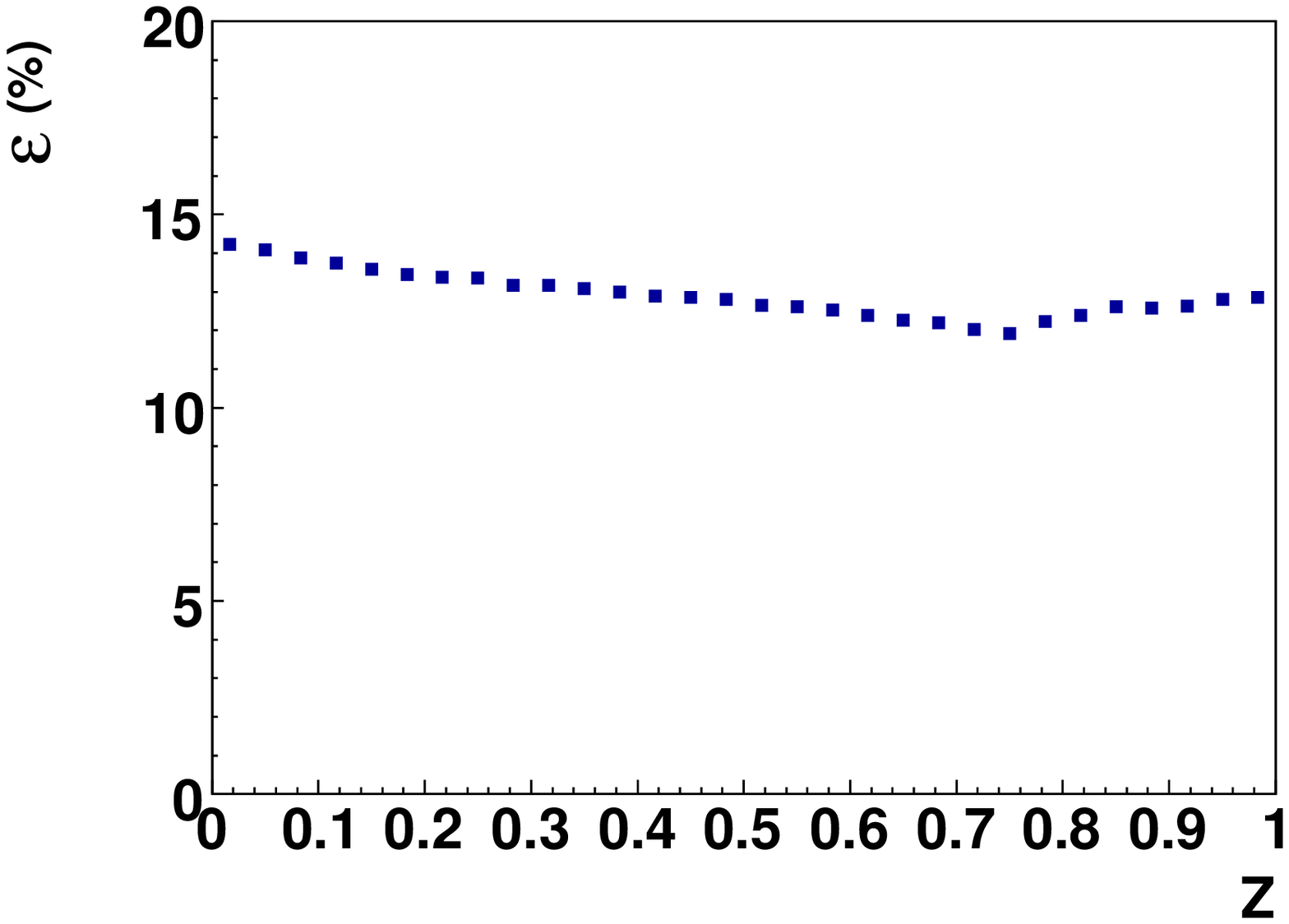}\\  
      \end{tabular}
      \caption{Medium purity sample. Left: Resolution on the $z$ variable. 
	Right: Reconstruction efficiency vs. $z$.} 
      \label{fig:res_eff} 
    \end{center}
  \end{figure}

\begin{figure}[!htb] 
    \begin{center}
     \includegraphics[width=.5\linewidth]{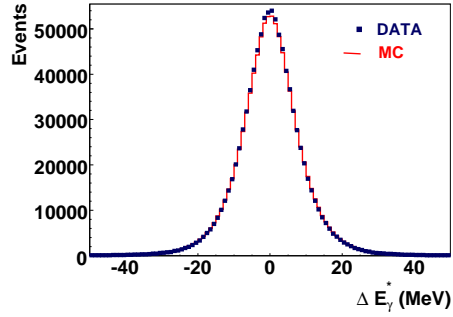}\\  
      \caption{Plot of $\Delta E^{*}_{\gamma} =
      E^{*}_{\gamma_{1}}-E^{*}_{\gamma_{2}}$, where $E^{*}_{\gamma}$
      are the $\gamma$ energies 
      from \piz decay in \piz CM. (Dots: data, histogram: MC).} 
      \label{fig:proj} 
    \end{center}
  \end{figure}

\begin{figure}[!htb] 
    \begin{center}
      \includegraphics[width=.5\linewidth]{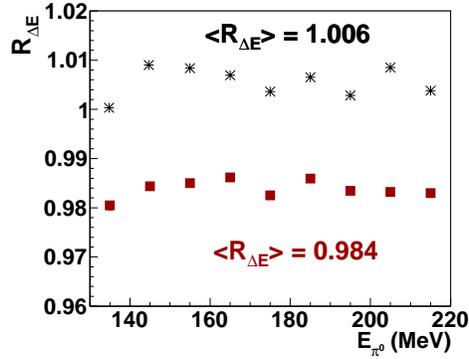}\\ 
    \end{center} 
    \caption{ $R_{\Delta E}$: ratio of $(\Delta E^{*}_{\gamma})^{data}_{rms} / (\Delta E^{*}_{\gamma})^{MC}_{rms} $  
      vs. $E_{\piz}$. Dots (stars) before (after) the correction for
      the difference between Data and MC.}
    \label{fig:rms} 
  \end{figure}

  The measurement is based on an integrated luminosity of  420 pb$^{-1}$ 
  corresponding to $\simeq 1.4 \cdot 10^{9}~\phi $ mesons produced.
  This data sample contains about 17 millions of $\eta$ mesons.\\
  The detector response to the decay of interest was studied by using
  the KLOE MonteCarlo (MC) simulation program~\cite{EVCL}. The MC
  takes into account variations in the machine operation and background
  conditions on a run-by-run basis.
  A MC sample for both signal and backgrounds was produced for an integrated luminosity five times that of the collected data.
  In the MC simulation of the $\eta\rightarrow 3 \piz$ decay, the
  signal has been generated using our preliminary
  measurement~\cite{K_p} of $\alpha = -0.027$.\\ 
  We search for: $\phi\rightarrow\eta\gamma$ with $\eta\rightarrow\piz\piz\piz$ and $\piz\rightarrow\gamma\gamma$ events. 
  To select the final state, we require to have seven prompt photons in the event.
  A photon is defined as an EMC cluster  not associated to a DC track. 
  We further require that $|(t-r/c)|<5\sigma_t$, where $t$ is the arrival time at the EMC, $r$ is the
  distance of the cluster from interaction point, IP, $c$ is speed of light.
  Fig.~\ref{fig:Egamma} shows the photon energy spectrum.
  The recoil photon from the two body decay $\phi\rightarrow\eta\gamma$
  is almost monochromatic, with $E_{\gamma\,rec} \simeq 363$ MeV and
  separated from the softer photons from $\piz$ decay.
  \\ 
  All events must pass a first-level selection to filter machine background 
  and an event classification procedure~\cite{EVCL}.
  Events with the expected final state signature are selected by requiring:
  \begin{itemize}
  \item $7$ and only $7$ prompt photons with $21^{\circ}<\theta_{\gamma}<159^{\circ}$ and $E_{\gamma}>10$ MeV.
    The angle between any photon pair, $\theta_{\gamma\gamma}$, must
    be $>9^{\circ}$ to reduce split showers.
    After these selection cuts we are left with $\simeq 4.6 \cdot
    10^{6}$ events.
  \item A constrained kinematic fit imposing total $4-$momentum conservation and $t = r / c$ for each photon is
    performed. Input variables to the fit are the energies, times
    of flight and the coordinates of clusters in the EMC and the beam
    energies. The fit improves the photon energies resolution: the
    $\pi^{0}$  mass resolution of $\sim 15.4$ MeV improves to $\sim 9.6$ MeV
    after applying the kinematic fit.  
    The selected events must satisfy the requirement $P_{\chi^{2}} >
    0.01$ corresponding to $\chi^{2}<25$.
    After this cut we are left with 1.9 millions of $\eta\rightarrow 3 \piz$ events corresponding to a
    signal efficiency of $( 40.81 \pm 0.01 ) \%$. At this level, the residual  
    background contamination, mainly due to $K_{S}K_{L}$ decays to neutral channels, is estimated by MC 
    to be $\sim 0.1$\%.
  \item To find the best combination (among 15) of the six less
    energetic photons into three $\piz$ a pairing procedure is
    applied. The procedure uses a pseudo-$\chi^{2}$ variable:
    \begin{equation}
      \chi^{2}_{j} = \sum_{i=1}^{3} \left( \frac{m_{\gamma\gamma ,i j } - m_{\pi^{0}}}
	  {\sigma_{m_{\pi^{0}}}}\right)^{2} \qquad\qquad j = 1,2,\ldots,15.
    \end{equation}
    \noindent 
    where $m_{\gamma\gamma ,i j }$ is the invariant mass of the
    $i^{th}$ photon pair, in corrispondence of the $j^{th}$ combination; 
    $\sigma_{m_{\pi^{0}}}$ is the corresponding $\pi^{0}$ mass resolution parametrized, 
    as function of the photon energy resolution: 
    \begin{equation}
      \frac{\sigma_{m_{\pi^{0}}}}{m_{\pi^{0}}} =
      \frac{1}{2}\left(\frac{\sigma_{E_{\gamma 1}}}{E_{\gamma 1}} \oplus
      \frac{\sigma_{E_{\gamma 2}}}{E_{\gamma 2}}\right), 
    \end{equation}
    the angular resolution contribution is negligible.
    In Fig.~\ref{fig:chi2_0} a data-MC comparison of the minimum value of the
    pseudo-$\chi^{2}$, $\chi^{2}_{min}$ is shown.
    \begin{figure}[!htb] 
      \begin{center}
	\includegraphics[width=.5\linewidth]{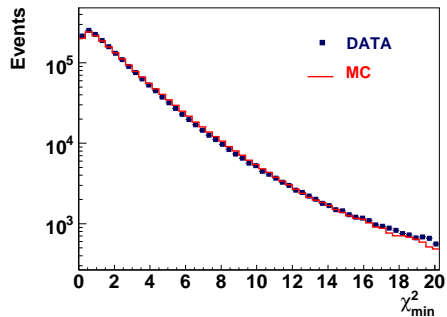}\\ 
	\caption{ Distribution of $\chi^{2}_{min}$, used to pair photons.(Dots: data, histogram: MC) } 
	\label{fig:chi2_0} 
      \end{center}
    \end{figure}\\ 
    The fraction of events with correctly paired photons, estimated from
    MC, is named in the following as purity, P, of the data
    sample. While we define WPf = 1 - P the wrong
    pairing fraction to $\pi^{0}$'s. To
    improve the purity a further cut is applied: $\chi^{2}_{min} < 5$.
    The distribution of
    the invariant mass of the two photons from $\piz$ decay,
    is shown in Fig.~\ref{fig:mass_gg}. 
  \begin{figure}[!htb] 
    \begin{center}
      \includegraphics[width=.5\linewidth]{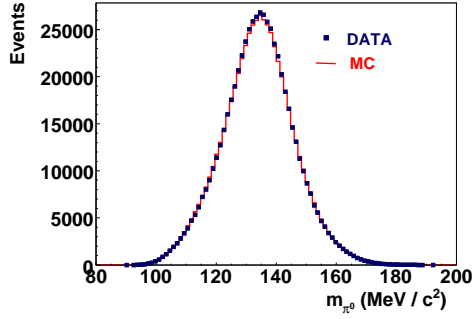}\\ 
    \end{center} 
    \caption{ Invariant mass of the two photons from $\piz$ decay after cut $\chi^{2}_{min} < 5$. (Dots: data, histogram: MC).}
    \label{fig:mass_gg} 
  \end{figure} 
\item After the photons pairing
    procedure a second kinematic fit is performed where the constraints on $\pi^{0}$
    and $\eta$ mass are also imposed. For the $\eta$ mass
    we used the value $547.874 \pm 0.007\,stat\; \pm 0.031\,syst\;$MeV measured by our experiment~\cite{Biagio}. 
    This fit improves the $z$ resolution by a factor two.  
  \end{itemize}
  \noindent
  We define three samples with different purity applying
  different cuts on the difference of the two lowest values of
  $\chi^{2}$, $\Delta \chi^{2}$, as reported in Table~\ref{tab:3_samples}.
  The resolution and efficiency as function of $z$ are shown in
  Fig.~\ref{fig:res_eff} for the Medium purity sample.
  The reconstruction efficiency, $\varepsilon(z)$, is obtained by MC for each $z$ bin, as the ratio: $\varepsilon(z) =
  N_{rec}(z) / N_{gen}(z)$ where $N_{gen,rec}$ are respectively the 
  generated and reconstructed events.\\
  \\ 
  \begin{table}[htb]
    \begin{center}
      \begin{tabular}{||c||c||c||c||c||} 
	\hline
	  {\bf$\Delta \chi^{2}$ cut } &   {\bf Samples}	& {\bf Purity}
	  & {\bf Efficiency} & {\bf N. events }\\ 
	    \hline      
	   2.5  & Low          & 90.4 \% & (20.07 $\pm$ 0.01)\% & 948471 \\
	    5   & Medium       & 95.0 \% & (12.96 $\pm$ 0.01) \% & 614663 \\
	    9   & High         & 97.3 \% &  (7.04  $\pm$ 0.01)\%  & 333493 \\
	\hline
      \end{tabular}
    \end{center}
    \caption{The three samples of different purity selected by
      different cuts on $\Delta {\chi^{2}}$.}
    \label{tab:3_samples}
  \end{table}
  The photon energy resolution is compared between data and MC
  looking at the distribution of $\Delta E^{*}_{\gamma} = E^{*}_{\gamma_{1}}-E^{*}_{\gamma_{2}}$; 
  i.e. the difference between photons energy in the $\piz$ rest frame.
  In Fig.~\ref{fig:proj} the distribution of $\Delta E^{*}_{\gamma}$ is plotted.
  Estimating the r.m.s. 
  of the $\Delta E^{*}_{\gamma}$ for
  slices of 10 MeV in $E_{\piz}$ a difference of ( 1 $\div$ 1.5)\% between data and MC is observed.
  Consequently, the MC photon energies
  have been smeared by this amount.\\
  Fig.~\ref{fig:rms} shows the ratio $R_{\Delta E \gamma} =(\Delta
  E^{*}_{\gamma})^{data}_{rms} / (\Delta E^{*}_{\gamma})^{MC}_{rms} $. 
  The correction improves the agreement between data and MC 
  on this variable.
  The residual difference, of $(0.6 \pm 0.2)$\%, is taken into account directly in the 
  evaluation of the result.
  
  \section{Measurement of the slope parameter $\alpha$}
  \label{section:fit}
  The fit to the Dalitz plot is done minimizing a log--Likelihood
  function built as follows:
  \begin{equation}
    -\log{\cal L}\left(\alpha\right) = -\sum_{i=1}^{N_{bin}} n_{i}
    \log\nu_{i}\left(\alpha\right),
    \label{eq:log_mult} 
  \end{equation}
  where, for each bin: $n_{i}$ are the number of reconstructed events,
  $\nu_{i}$ the number of expected events, obtained from MC taking into account the detector
  resolution and WPf and weighted with
  $1 + 2\alpha z$. Moreover we
  correct for the data-MC differences in the WPf. To estimate it on data
  we use the distribution of $z$ variable reconstructed using the second best pairing combination,$z_{\chi^{2}_{2}}$.
  This distribution is fit with the superposition of the
  MC shapes for events with good and wrong pairing respectively.
  The uncertainty on the WPf data-MC difference is taken into account in
  evaluating the systematic error.

\begin{figure}[!htb] 
    \begin{center}
      \begin{tabular}{c c}
	\includegraphics[width=.5\linewidth]{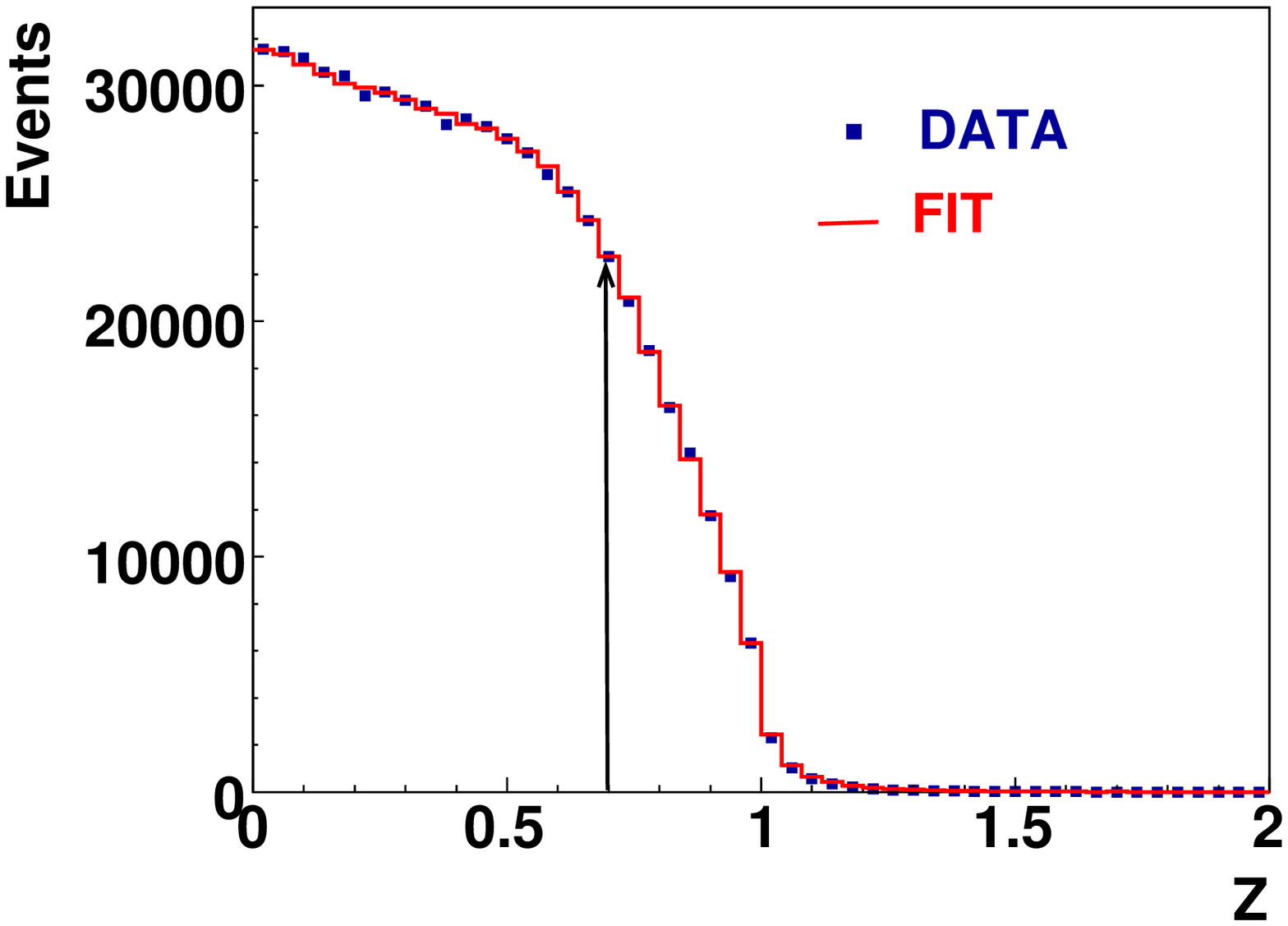}&
	\includegraphics[width=.5\linewidth]{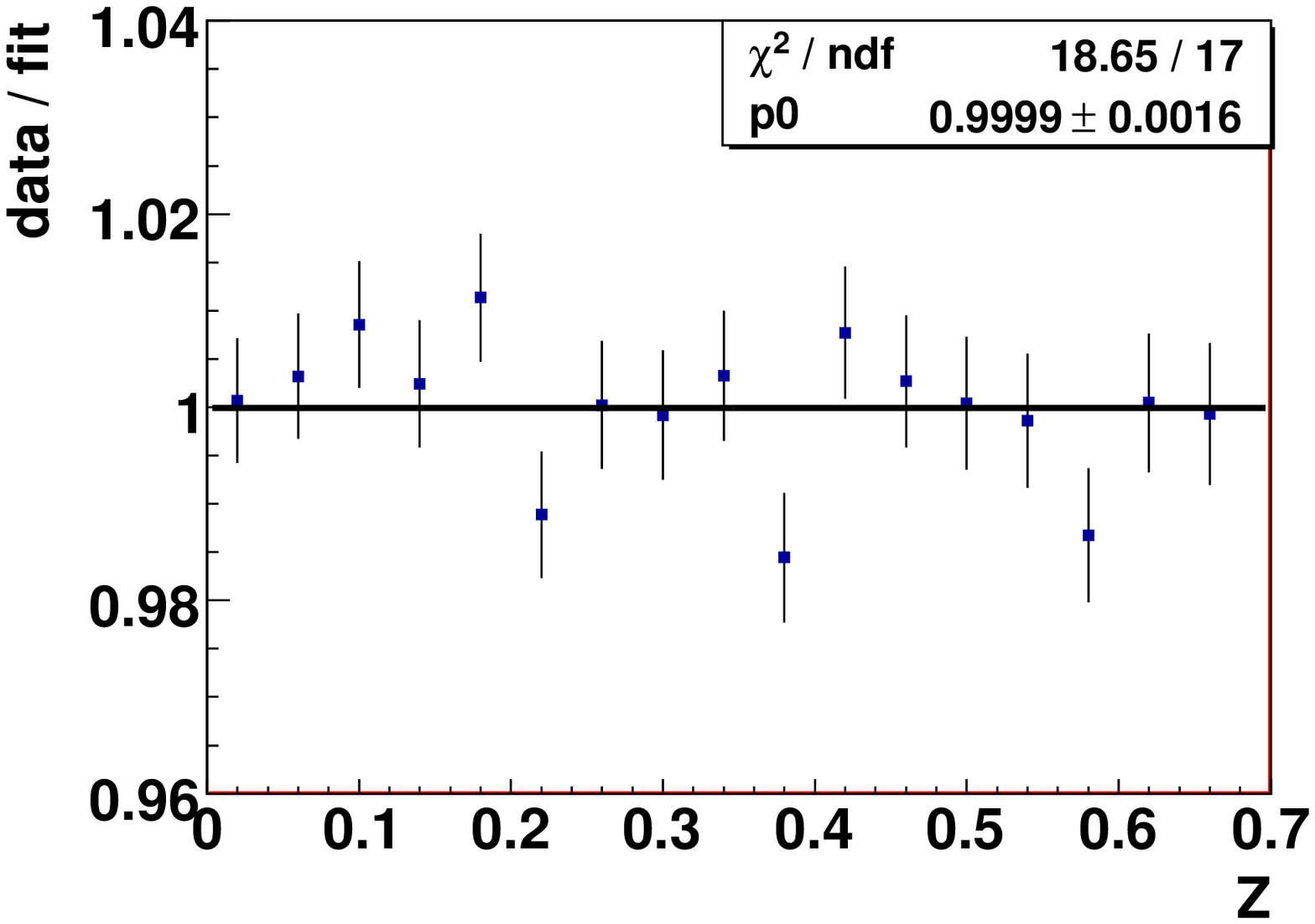}\\  
    \end{tabular}   
    \end{center} 
    \caption{Medium purity sample: Left: observed $z$ distribution with the
    corresponding fit overimposed. Right: data/fit ratio as function
    of $z$. }
    \label{fig:fit_data} 
  \end{figure}

  The fit  procedure has been tested on MC by verifying that the
 fit reproduces in output the same input value, within the statistical
 error.\\
  To obtain the final result the
  fit range $(0 \div 0.7)$, corresponding to the region of the phase
  space in which the $z$ distribution is flat, and the 
  Medium purity sample is chosen.
  The fit results for the three different Purity samples are shown in Table ~\ref{tab:resfit_1}. Moreover, we have
  applied a shift of $\Delta \alpha = -0.0008$ on the slope parameter
 $\alpha $ to correct the residual data-MC
 discrepancy in the photons energy resolution, see Section~\ref{Section:selezione}.   
\begin{table}[htb]
  \begin{center}
    \begin{tabular}{||c||c||c||c||}
      \hline
      {} &  
      {\small{\bf Low Purity   }} &  
      {\small{\bf Medium Purity  }}&
      {\small{\bf High Purity  }}\\ \hline
      ${\bf \alpha \cdot 10^{4}}$  &$-319\pm 29$ & $-301\pm 35$  &$-308\pm 47$\\
      \hline     
      ${\bf P_{\chi^{2}}}$  &$92$\%  & $85$\%  &$91$\% \\
      \hline
    \end{tabular}
  \end{center}
  \caption{$\alpha$ values from fit for different purity data
    samples.}
  \label{tab:resfit_1}
\end{table}\\
In Fig.~\ref{fig:fit_data} 
a comparison between the observed and fitted $z$ distributions
is shown.
  
\section{Systematic uncertainties}
\label{sistematiche}
In the following we describe the sources of systematics. For each
of them, the fit has been repeated varying the related sources and assuming as systematic
error the difference with
respect to the reference value. In Table ~\ref{tab:syst} we have summarized
all the systematic errors.
\begin{itemize}
\item {\bf Analysis cuts}
  To  control the stability of the result respect to our analysis cuts
  we have moved them independently. The cut on
  $\theta_{\gamma\gamma}$ to reject split showers,, was varied in the 
  interval $6^{\circ} - 18^{\circ}$ in steps of $3^{\circ}$.
  The photon energy threshold was also
  increased from 10 MeV to 40 MeV with a step of 5 MeV. The related
  systematic error is very small.
\item {\bf Energy Resolution} 
  As shown in Fig.~\ref{fig:rms}, the data-MC comparison of $(\Delta
  E^{*}_{\gamma})_{rms}$  after correction shows a residual
  discrepancy of $(0.6 \pm 0.2)$\%. While the 0.6\% correction has already been applied, we estimate
  the systematics related to its uncertainty to be $\Delta \alpha = \pm 3 \cdot 10^{-4}.$
\item {\bf $\eta$ mass} 
  This systematic effect has been estimated
  varying the $\eta$ mass on data by $\pm 0.031$ MeV accordingly to
   our measurement~\cite{Biagio}.
\item {\bf Wrong pairing fraction}
  For the sample used the data-MC ratio of
  WPf is 1.1 $\pm$ 0.1.
  As mentioned in Section~\ref{section:fit} the fit procedure takes 
  into account this difference. To assign the systematic error we
  repeated the fit procedure varying the WPf within the $\pm 10$\%
  uncertainty quoted above.
\item {\bf Purity} 
  As a check of the MC capability to reproduce the samples purity and
  its dependence upon $z$, we show in Fig.~\ref{fig:Nhigh_low} the ratio
  between the number of events for the High and the Low purity sample,
  $N_{High}/N_{Low}$, as a function of $z$. A good agreement
  between data and MC throughout the fit range is observed. 
 As systematic error, we take the
  difference between the $\alpha$ values estimated using the Low and the High
purity sample, see Table~\ref{tab:resfit_1}. 
  \begin{figure}[!htb] 
    \begin{center}
      \includegraphics[width=.5\linewidth]{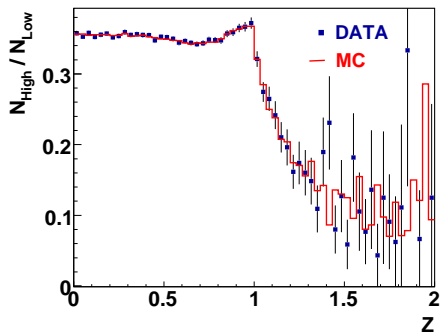}\\  
    \end{center} 
    \caption{Ratio $N_{high}/N_{low}$ as a function of $z$. (Dots: data, histogram: MC).  }
    \label{fig:Nhigh_low} 
  \end{figure} 
\item {\bf Fit range and binning} 
  The fit was repeated with different values of
  the fit range· from $[0 \div 0.6 ]$ to $[0 \div 1 ]$ with a step of 0.1. This is
  the largest systematic effect. Instead we find negligible effect when
  changing the bin size by a factor 2 from 0.04
  to 0.02.
\end{itemize}

\begin{table}[htb]
  \begin{center}
    \begin{tabular}{||c||c||}
      \hline
	  {\small{\bf Source}} &  
	  {\small{\bf $\Delta \alpha  \cdot 10^{4}$ }} \\  
	  \hline \hline
	  Analysis cuts                 & -1 $\;\;\;$ +1 \\
	  Energy resolution             & -3 $\;\;\;$ +3 \\
	  $\eta$ Mass                   & -2 $\;\;\;$ +6 \\
	  Wrong pairing                 & -6 $\;\;\;$  +5 \\
	  Purity                        & -18 $\;\;\;$  +0    \\      
	  Fit range                     & -29 $\;\;\;$ +20\\
	  \hline	
	  Total                         & -35 $\;\;\;$  +22\\	  
	  \hline
    \end{tabular}
  \end{center}
  \caption{ Summary of the systematic errors on the slope parameter
    $\alpha$. The total systematic error is the sum in quadrature of the
    different contributions.}
  \label{tab:syst}
\end{table}
\section{Conclusions}
Using a clean sample of $\eta\rightarrow 3\pi^{0} $ decays we have
measured the Dalitz Plot slope parameter obtaining
$\alpha = -0.0301 \pm 0.0035\,stat\;_{-0.0035}^{+0.0022}\,syst\,$ in
agreement with other recent results of comparable precision.\\
The above value is also consistent with
$\alpha = -0.038 \pm 0.003\,stat\;_{-0.008}^{+0.012}\,syst\,$
obtained from the KLOE study of the \etapippimpiz decay~\cite{NOI}
using the theoretical correlations between the two decay modes.\\
Our $\alpha$ measurement confirms the inadequacy of simple
NLO ChPT  computations and the need to take into account higher order
corrections.

\section*{Acknowledgements}
We thank the DAFNE team for their efforts in maintaining low background running 
conditions and their collaboration during all data-taking. 
We want to thank our technical staff: 
G.F.Fortugno and F.Sborzacchi for their dedicated work to ensure an
efficient operation of 
the KLOE computing facilities; 
M.Anelli for his continuous support to the gas system and the safety of
the
detector; 
A.Balla, M.Gatta, G.Corradi and G.Papalino for the maintenance of the
electronics;
M.Santoni, G.Paoluzzi and R.Rosellini for the general support to the
detector; 
C.Piscitelli for his help during major maintenance periods.
This work was supported in part
by EURODAPHNE, contract FMRX-CT98-0169; 
by the German Federal Ministry of Education and Research (BMBF) contract 06-KA-957; 
by the German Research Foundation (DFG),'Emmy Noether Programme',
contracts DE839/1-4;
and by the EU Integrated
Infrastructure
Initiative HadronPhysics Project under contract number
RII3-CT-2004-506078.


\end{document}